\documentclass[twocolumn,epjc3]{svjour3}          

\RequirePackage{mathptmx}      
\usepackage{amsmath}
\usepackage{amssymb}
\usepackage{latexsym}
\usepackage{amsmath}
\RequirePackage{flushend}
\RequirePackage[numbers,sort&compress]{natbib}

\journalname{Eur. Phys. J. C}
\begin{document}

\title{Novel discrete symmetries in the general ${\cal N} = 2$ supersymmetric quantum mechanical model}


\author{R. Kumar\thanksref{e1, addr1}
        \and
        R. P. Malik\thanksref{e2, addr1}
}

\thankstext{e1}{e-mail: raviphynuc@gmail.com}
\thankstext{e2}{e-mail: rpmalik1995@gmail.com}


\institute{Department of Physics, Centre of Advanced Studies, Faculty of Science,
Banaras Hindu University, \\Varanasi - 221 005, U. P., India \label{addr1}
           \and
DST Centre for Interdisciplinary Mathematical Sciences,
Faculty of Science, Banaras Hindu University, \\Varanasi - 221 005, U. P., India\label{addr2}
           }

\date{Received: date / Accepted: date}

\maketitle

\begin{abstract}
In addition to the usual supersymmetric (SUSY) continuous symmetry transformations for the 
general ${\cal N}= 2$ SUSY quantum mechanical model, we show the existence of 
a set of novel discrete symmetry transformations for the Lagrangian  of the above SUSY 
quantum mechanical model. Out of all these discrete symmetry transformations, a unique discrete 
transformation corresponds to the Hodge duality operation of differential 
geometry and the above SUSY continuous symmetry transformations 
(and their anticommutator) provide the
physical realizations of the 
de Rham cohomological operators of differential geometry. Thus, we provide a concrete proof of  our 
earlier conjecture that any arbitrary ${\cal N}= 2$ SUSY quantum mechanical 
model is an example of a Hodge theory where the cohomological operators find their 
physical realizations in the language of symmetry transformations of this theory.  
Possible physical implications of our present 
study are pointed out, too.\\

\end{abstract}

\section{Introduction}
The supersymmetric (SUSY) models of quantum mechanics (QM) provide one of the most fertile 
grounds for the growth of ideas, germinating out from the branches of  mathematics and physics,  in a physically  
meaningful manner (see, e.g. \cite{1,2,3}). A decisive feature of these models is the observation that, 
at the classical level, the commuting variables of the theory transform to their anticommuting 
counterparts and vice-versa due to the presence of a fermionic SUSY symmetry in the theory. At the quantum 
level, SUSY QM admits two Hamiltonians (and corresponding states) which are connected to
each-other, in a specific manner, because of the presence of the above quoted fermionic SUSY symmetry.
The existence of the latter is one of the hallmarks of  any arbitrary SUSY quantum mechanical 
theory in any arbitrary  dimension of spacetime.

A very special class of the above SUSY QM is the one  which is characterized 
by the existence of  two fermionic $(Q^2 = \bar Q^2 = 0)$  charges ($Q$ and $\bar Q$) 
and  the (bosonic) Hamiltonian $(H)$ of 
the theory which obey a specific ${\cal N} = 2$ SUSY algebra. 
All these  charges generate continuous symmetry transformations 
that also satisfy the above algebra in their operator form. These sets of SUSY 
quantum mechanical systems are known as the general ${\cal N} = 2$ SUSY models. 
Some of the physical examples of such a class of theories  have 
been recently shown to be endowed with a novel set of discrete symmetry transformations. The interplay 
of the discrete and usual continuous symmetry transformations has led to establish that the
${\cal N} = 2$ SUSY models   provide the {\it physical} examples of Hodge theory \cite{4,5}.

In a very recent paper \cite{5}, we also conjectured that any arbitrary ${\cal N} = 2$ 
SUSY quantum mechanical model would provide a physical example for the Hodge theory 
where the  de Rham cohomological operators, Hodge duality operation, degree of a form, etc., 
of differential geometry would find their physical realizations in the language of 
symmetry properties of the above SUSY systems. The purpose of our present investigation 
is to give the proof of the above conjecture for any arbitrary ${\cal N} = 2$ SUSY model 
with any arbitrary superpotential. We show that this general model is endowed with 
discrete symmetry transformations which, together with the usual continuous symmetry 
transformations, provide the physical realizations of {\it all} the cohomological 
operators of differential geometry. In fact, there exists a novel duality symmetry in 
the theory and all aspects of the cohomological operators are realized in the language 
of symmetry properties and  conserved charges (and their eigenvalues).

In our earlier works \cite{6,7,8,9,10}, we have shown that the 2D usual (non-)Abelian 1-form gauge 
theories, 4D Abelian 2-form gauge theory and 6D Abelian 3-form gauge theory provide 
perfect models for the Hodge theory within the framework of Becchi-Rouet-Stora-Tyutin (BRST) formalism.
Furthermore, exploiting the Hodge decomposition 
theorem and choosing the physical state to  be the harmonic state, we have shown that the 2D 
(non-)Abelian 1-form gauge theories  provide a new model for the topological field theory 
(TFT) \cite{6,11} which captures a part of the key aspects of Witten-type TFTs \cite{12} and 
some salient  features of Schwarz-type TFTs \cite{13}. None of the above theories 
 \cite{6,7,8,9,10,11} are, however, supersymmetric in nature. The central goal of our present endeavour  is to first 
study the general ${\cal N} = 2$ SUSY quantum mechanical system and 
gain deep insights into its mathematical and physical structures and comment on the generalization 
of these ideas, if possible, to the study of ${\cal N} = 2$ SUSY gauge theories of phenomenological interest.

The following factors have propelled us to pursue our present investigation. 
First and foremost, it is of utmost importance for us to provide a concrete 
proof of our earlier conjecture that any arbitrary ${\cal N} = 2$ SUSY quantum 
mechanical model would provide a tractable physical example of a Hodge theory. Second, 
it is always an important endeavour to state a general rule for the solution of 
a given problem. In our present work, we have provided a general answer to a 
general question of physical importance. Finally, we are very hopeful that we shall 
be able to apply our current ideas to ${\cal N} = 2$ SUSY gauge theories  which might 
turn out to be the field theoretic model  for a (quasi-)topological field theory as well as an 
example of a Hodge theory.

In the broader  perspective, our present study is essential besides the above cited factors
of motivations. It is quite possible that our present study would
enable us to count the correct degrees of freedom associated with the  
fields of a ${\cal N} = 2$ SUSY quantum gauge theory in a given dimension of spacetime. To achieve the above goal,
we shall have to exploit the Hodge decomposition theorem in the quantum Hilbert space of this specific
SUSY gauge theory under consideration.  Since the most symmetric state would turn out to be the 
harmonic state, we shall be forced to choose it as the physical state of the theory.
The annihilation of this state by the supercharges would lead to
the correct counting of the degrees of freedom.  It is gratifying to state that we have already performed such
kind of analysis in our earlier works on the {\it usual} 2D free (non-)Abelian gauge
theories \cite{6,11}. We are very hopeful that our ideas of the earlier works  \cite{6,7,8,9,10,11}
would persist even in the case of ${\cal N} = 2$ SUSY quantum gauge theories in a specific dimension of spacetime
where we shall be able to obtain the perfect SUSY quantum gauge models 
for the Hodge theory as well as, possibly, a set of SUSY examples of TFTs.

The contents of our present paper are organized as follows. In Sect. 2, we concisely 
recapitulate the bare essentials of superspace approach to ${\cal N} = 2$ SUSY QM 
and derive the Lagrangian and its associated  continuous SUSY transformations. Our Sect. 3 deals 
with the derivation of conserved charges from the Noether theorem. We discuss various 
discrete symmetry transformations of the Lagrangian of our theory in Sect. 4.  
We devote time on the algebraic structures of various symmetry transformations in  Sect. 5 and show their 
relevance to the cohomological  aspects of differential geometry. We establish mapping between 
the  conserved charges and cohomological operators in our Sect. 6. Finally, we make a 
few  concluding remarks in Sect. 7.

In our Appendix, we provide the proof for the specific ${\cal N} = 2$ algebra amongst the 
supercharges and Hamiltonian of our present  SUSY theory in a simpler form.

\section{Preliminaries: superspace approach to the description of ${\cal N} = 2$ SUSY quantum mechanical model}

We begin with a ${\cal N} = 2$ supervariable $X(t, \theta, \bar \theta)$ on a $(1, 2)$-dimensional supermanifold 
which is characterized by the superspace coordinates $Z^M = (t, \theta, \bar \theta)$ where $t$ 
is the bosonic ($t^2 \ne 0$) variable  and $(\theta, \bar \theta)$ 
are a pair of Grassmann variables 
(with $\theta^2 = \bar \theta^2 = 0,\; \theta\, \bar \theta + \bar \theta\, \theta = 0$).
We note that, later on, the ordinary variable $t$ would turn out to be the  evolution 
parameter for the SUSY quantum mechanical system. 
The above supervariable can be expanded, along the Grassmannian directions, as \cite{14,15}
\begin{eqnarray}
X(t, \theta, \bar \theta) = x(t) + i \,\theta\,\bar \psi(t) 
+ i \,\bar \theta\, \psi (t) + \theta\, \bar \theta \,A(t),
\end{eqnarray}
where ($x(t),\; \psi (t),\; \bar \psi (t)$) are the  basic dynamical  variables and $A(t)$ 
is an auxiliary variable for our SUSY quantum mechanical system. We note that the variables ($x(t),\; A (t)$) 
form a bosonic pair and ($\psi (t),\; \bar \psi (t)$) (with $\psi^2 = \bar \psi^2 = 0, \; \psi\,\bar\psi 
+ \bar \psi\, \psi = 0$)  are their SUSY  counterparts (i.e. fermionic pair of variables). 
All these variables are function of the evolution parameter $t$.

The two supercharges $Q$ and $\bar Q$ (with $Q^2 = 0, \; \bar Q^2 = 0$), for 
the above ${\cal N} = 2$ SUSY general quantum mechanical theory, are defined as \cite{14,15}
\begin{eqnarray}
Q = \frac{\partial}{\partial \bar \theta} + i\,\theta\,\frac{\partial}{\partial t}, \qquad \qquad
\bar Q = \frac{\partial}{\partial  \theta} + i\, \bar \theta\,\frac{\partial}{\partial t},
\end{eqnarray}
where $\partial_M = \partial/\partial Z^M
= (\partial/\partial t,\; \partial/\partial \theta,\; \partial/\partial \bar \theta)$
are the partial derivatives defined on the $(1, 2)$-dimensional supermanifold. 
The latter turn out to be the generators for  the translations   along $(t, \theta, \bar \theta)$ directions 
as illustrated below:
\begin{eqnarray}
&&t\longrightarrow t' = t + i \,(\epsilon\, \bar \theta + \bar \epsilon\, \theta), \qquad
\theta \longrightarrow \theta' = \theta + \epsilon, \nonumber\\
&& \bar \theta \longrightarrow \bar \theta' = \bar \theta + \bar \epsilon, 
\end{eqnarray}
where $\epsilon$ and $\bar \epsilon$ are the infinitesimal shift 
transformation parameters along the Grassmannian directions of the $(1, 2)$-dimensional supermanifold. 
Thus, they are fermionic ($\epsilon^2 = 0, \; \bar \epsilon^2 = 0, \; \epsilon\, \bar \epsilon 
+ \bar \epsilon\, \epsilon = 0$) in nature.

The SUSY transformation $(\delta)$ on the supervariable  
\begin{eqnarray}
\delta X(t, \theta, \bar \theta) &=& \delta x(t) + i\, \theta\, \delta\bar \psi(t) + i \,\theta\, \delta \psi(t) 
+ \theta\, \bar \theta\, \delta A(t) \nonumber\\
&\equiv& (\bar \epsilon\, Q + \epsilon \, \bar Q)\, X(t, \theta, \bar \theta),
\end{eqnarray}
can be expressed in terms of the supercharges $Q$ and $\bar Q$ as illustrated above.
The transformation $(\delta)$ can be divided into two infinitesimal 
transformations $\delta_1$ and $\delta_2$ because of the presence of ${\cal N} = 2$ SUSY QM.
These are juxtaposed as: 
\begin{eqnarray}
&& \delta_1 x = i\, \bar \epsilon\,  \psi,   	\hskip 2.5cm \delta_2 x = i\, \epsilon \, \bar \psi,  \nonumber\\
&& \delta_1 \bar \psi = - \bar \epsilon\,(\dot x + i\, A), \hskip 1.4 cm \delta_2  \psi = -  \epsilon\, (\dot x - i\, A),\nonumber\\
&& \delta_1 A =  - \bar \epsilon\,\dot \psi,	 \hskip 2.3cm \delta_2 A = \epsilon\,\dot {\bar \psi},\nonumber\\
&& \delta_1 \psi = 0,						 \hskip 2.8cm \delta_2 \bar  \psi = 0,
\end{eqnarray}
where we have defined $\dot x = dx/dt,\; \dot \psi = d\psi/dt,\; \dot {\bar \psi} = d \bar \psi/dt.$
It can be readily checked that $\delta_1$ and $\delta_2$ are off-shell nilpotent of order two 
(i.e. $\delta^2_1 = 0, \; \delta^2_2 = 0$).

The general Lagrangian for the ${\cal N} = 2$ SUSY quantum mechanical model can be written, in terms of ${\cal D}, 
\bar {\cal D}$ and $W$, as (see, e.g. \cite{14,15} for details)
\begin{eqnarray}
L &=& \int d\theta \,d\bar \theta \bigg[\frac{1}{2}\,{\cal D} X(t, \theta, \bar \theta)\, 
\bar {\cal D} X(t, \theta, \bar \theta) \nonumber\\
&-& W\left(X(t, \theta, \bar \theta)\right)\bigg],
\end{eqnarray}
where ${\cal D}$ and $\bar{\cal D}$ are the following supercovariant derivatives
\begin{eqnarray}
{\cal D} = \frac{\partial}{\partial \bar \theta} - i\,\theta\,\frac{\partial}{\partial t}, \qquad
\bar {\cal D} = \frac{\partial}{\partial  \theta} - i\, \bar \theta\,\frac{\partial}{\partial t},
\end{eqnarray}
and $W (X)$ is the superpotential which  is an arbitrary function of the supervariable $X(t, \theta, \bar \theta)$.
One can expand the above superpotential $W(X (t, \theta, \bar\theta))$, by 
Taylor expansion, around the ordinary space variable 
$x$, as follows:  
\begin{eqnarray}
W\big(X(t, \theta, \bar \theta)\big) &=& W(x + i\,\theta \, \bar \psi 
+ i\, \bar \theta \, \psi + \theta\, \bar \theta\, A)\nonumber\\
&=& W(x) + (i\,\theta \, \bar\psi + i\, \bar \theta \, \psi + \theta\, \bar \theta\, A)\, W'(x)\nonumber\\ 
&+& \frac{1}{2!}\, (i\,\theta \, \bar\psi + i\, \bar \theta \, \psi + \theta\, \bar \theta\, A)^2\,W''(x).
\end{eqnarray} 
We note that there will be no further higher order terms in the above expansion.
Finally, performing the proper Grassmannian integration in (6),  we obtain the following physical 
Lagrangian for the ${\cal N} = 2$ SUSY model: 
\begin{eqnarray}
L &=& \frac{1}{2}\,{\dot x}^2 + \frac{i}{2}\,\Big(\bar \psi (t) \, \dot \psi (t) 
- \dot {\bar \psi} (t) \, \psi (t) \Big) + A(t)\, W'(x) \nonumber\\
&+& \frac{1}{2}\, A^2(t) +  \frac{1}{2}\, \Big(\bar \psi (t) \, \psi (t) 
- \bar \psi (t) \, \psi (t) \Big) W''(x),
\end{eqnarray}
where $W(x)$ is an arbitrary  superpotential and its first- and second-order derivatives
are: $W'(x) = \frac{d}{dx}W(x),$ $\; W''(x) = \frac{d^2}{dx}W(x)$. 
The above Lagrangian can be simplified, modulo a total derivative, as:
\begin{eqnarray}
L &=& \frac{1}{2}\,{\dot x}^2 + i\,\bar \psi (t) \, \dot \psi (t) 
 +  W'(x) \, A(t) + \frac{1}{2}\, A^2(t) \nonumber\\
&+&  W''(x)\,\bar \psi (t) \, \psi (t),
\end{eqnarray}
because of the fact that $\psi\, \bar \psi + \bar \psi\, \psi = 0$.  
We shall focus on the above Lagrangian (10) for our further discussions (in the rest of our discussions).

\section{Continuous symmetries: conserved charges}
The continuous SUSY transformations  $\delta_1$ and $\delta_2$ 
(cf. (5)) can be re-expressed in terms of the fermionic 
$(s^2_1 = 0, \; s^2_2 = 0)$ symmetry transformations if we identify $\delta_1 = \bar \epsilon \,s_1$ 
and $\delta_2 = \epsilon\, s_2.$ In explicit form, these transformations are
\begin{eqnarray}
&& s_1 x = i\, \psi, \;\quad \qquad\qquad s_2 x = i\, \bar \psi,\nonumber\\
&& s_1 \bar \psi = - (\dot x + i\, A),  \qquad s_2 \psi = - (\dot x - i\, A),\nonumber\\
&& s_1 A = - \dot \psi, \;\;\;\qquad\qquad  s_2 A =  \dot {\bar \psi},\nonumber\\
&&s_1  \psi = 0, \,\;\;\qquad\quad\qquad s_2 \bar \psi = 0.
\end{eqnarray}
We note that $s_1$ and $s_2$ are off-shell nilpotent of order two. In other words, we observe the validity of  
$s^2_1 = 0, \; s^2_2 = 0$ in their operator form where any equation of motion, emerging  from (10), is {\it not}
 used. The Lagrangian (10) transforms, under infinitesimal continuous transformations $s_1$ and $s_2$, as  
\begin{eqnarray}
s_1 L = \frac {d}{dt} \Big[- W'\,\psi\Big], \quad 
s_2 L = \frac{d}{dt}\Big[i\,\bar \psi\, \big(\dot x - i\,A\big) + \bar \psi\, W' \Big].
\end{eqnarray}
This observation establishes that the action integral $S = \int L \,dt$ remains invariant under the continuous
SUSY transformations $s_1$ and $s_2$.

One of the key ingredients of a ${\cal N} = 2$ SUSY theory is the fact that two successive SUSY 
transformations must generate the spacetime translation in a given dimension of spacetime. 
In our one-dimensional case, we have the following relationship for the generic variable $\Phi$, namely;
\begin{eqnarray}
&& s_w \Phi \equiv \{s_1,\; s_2\}\, \Phi = -2\,i\,\dot\Phi, \nonumber\\
&& \Phi = x,\;\psi, \;\bar \psi,\; A,\; W, \;W',\;W''.
\end{eqnarray}
Thus, modulo $(-2\,i)$ factor, we have the time translation for a variable if we apply the 
two successive SUSY continuous symmetry transformations. In other words, we have the 
following symmetry transformation:
\begin{eqnarray}
&&s_w L  = \big(s_1\,s_2 + s_2\,s_1\big)\,L \equiv \frac{d}{dt}\, L.
\end{eqnarray}
It is obvious, at this juncture,   that we have {\it three} standard continuous symmetry transformations in the theory. 
Two of them are fermionic $(s^2_1 = 0,\; s^2_2 = 0)$ and one of them $(s_w)$ is bosonic. The latter 
is obtained from the anticommutator between  the above fermionic symmetry transformations
(i.e. $s_w = \{s_1, \; s_2\}$).

By exploiting the standard techniques of Noether theorem, one can compute the 
conserved charges, corresponding to the above continuous symmetry transformations. The 
expressions, for the complete set of these charges, are
\begin{eqnarray}
 Q &=& (i\, \dot x - A)\,\psi, \qquad\qquad \bar Q = \bar\psi\,(i\, \dot x + A),  \nonumber\\
Q_w &=& \frac{1}{2}\, {\dot x}^2 - \frac{1}{2}\, A^2 - A\, W' - W''\, \bar \psi\, \psi \nonumber\\
&\equiv & \frac{1}{2}\, {p}^2 - \frac{1}{2}\, A^2 - A\, W' - W''\, \bar \psi\, \psi = H,
\end{eqnarray}
where $H$ is the Hamiltonian of the theory and $p = \dot x$ is the momentum w.r.t. $x$.
The conservation of the above charges (i.e. $\dot Q = 0,\; \dot {\bar Q} = 0, \; \dot Q_w = 0$) can be 
proven by exploiting the following Euler-Lagrange equations of motion:
\begin{eqnarray}
&&\ddot x - A\,W'' - W'''\, \bar \psi\, \psi =0, \qquad A = - W', \nonumber\\
&& \dot \psi - i\, W''\, \psi = 0 \Longrightarrow \ddot {\bar \psi} + i\, W'''\, \dot x\, \bar \psi 
+ (W'')^2 \,\bar \psi = 0,  \nonumber\\
&& \dot {\bar \psi} + i\, W''\, \bar \psi = 0 \Longrightarrow \ddot \psi - i\, W'''\, \dot x\,  \psi 
+ (W'')^2 \, \psi = 0,
\end{eqnarray}
which emerge from the Lagrangian (10) of the theory. The other way of proving the conservation 
law is to exploit the canonical (anti)commutators from the Lagrangian (10) and check 
that the commutators $[H,\; Q] = 0,\; 
[H, \; \bar Q] = 0$  and $[H, \; H] = 0$ are trivially satisfied. From the Heisenberg's 
equation of motion, this will,  ultimately, imply the conservation law 
($\dot Q = \dot{\bar Q} = \dot H = 0$).

\section{Discrete symmetries: duality transformations for the ${\cal N} = 2$ SUSY quantum mechanical model}

We focus here on a set of discrete symmetries of the Lagrangian (10) for our present general ${\cal N} = 2$
SUSY model of QM. We observe that under the following discrete transformations
\begin{eqnarray}
&& x \to -\,x, \qquad t\to +\, t, \qquad \psi(t) \to \pm \,i\,\bar \psi(t), \nonumber\\
&& \bar \psi(t) \to \mp\, i\, \psi(t),\qquad
 W'(x) \to + \, W'(x), \nonumber\\
&& W''(x) \to - \,W''(x),  \qquad A(t) \to +\, A(t), 
\end{eqnarray}
the Lagrangian (10) remains invariant. It is to be noted that, in the above, we have primarily 
{\it two} discrete symmetry transformations for the Lagrangian (10) where $\psi \to \pm\,i\, \bar \psi$ means 
$\psi(t) \to \psi'(t) = \pm\, i\,\bar \psi(t)$ and $W'(x) \to +\, W'(x)$ explicitly implies 
$W'(x) \to W'(-x) = +\, W'(x)$. As a consequence, the first derivative on  
 the potential function $W'(x)$ is {\it even} under the parity transformation 
(i.e. $\hat P \, W'(x) = W'(-x) \equiv +\, W'(x)$). We note further that there is a parity symmetry in the theory
but there is {\it no} non-trivial time-reversal symmetry (as $t \to t$).
It is to be emphasized that here the prime on $\psi(t)$ does not mean the space
derivative. Rather, the prime here  corresponds to an internal discrete transformation like: 
$\psi(t) \to \psi' (t) = e^{\pm\, i \pi/2}$ $\bar \psi (t)$. We repeat  that there are two hidden symmetries 
($ \psi \to \pm \,i\, \bar\psi$) in the above discrete symmetry transformations.

One can invoke the above cited time-reversal symmetry in the theory by 
checking that the following discrete transformations:
\begin{eqnarray}
&& x \to -\,x,\qquad  t\to  -\, t, \qquad \psi(t) \to \pm\,i \,\bar \psi(t), \nonumber\\
&& \bar \psi(t) \to \pm\, i \,\psi(t), \qquad W'(x) \to -\,  W'(x), \nonumber\\
&& W''(x) \to +\, W''(x),  \qquad  A(t) \to  -\,A(t),
\end{eqnarray} 
also leave  the Lagrangian invariant (i.e. $L \to L$). We explicitly mean, 
by the above transformations, the  following   
\begin{eqnarray}
\hat P&:& x \to - \,x, \qquad \hat P\, W'(x) \equiv W'(- x) = - \,W'(x),\nonumber\\
\hat t&:& t \to - \,t, \qquad \hat T \,\psi(t) \equiv \psi(- t) = \pm \,i  \bar\psi(t).
\end{eqnarray} 
Similarly, the other transformations can be interpreted. Thus, we figure out that there are parity 
and time-reversal symmetries {\it together} in (18).

There is yet another discrete symmetry in the theory which, as we shall see later on, 
plays a crucial role in our further discussions. The following discrete symmetry transformations 
\begin{eqnarray}
&& x \to -\,x, \qquad t\to  -\, t, \qquad \psi(t) \to +\,\pm\,\bar \psi(t), \nonumber\\
&& \bar \psi(t) \to \mp\,\, \psi(t), \qquad
 W'(x) \to -\,  W'(x),\nonumber\\
&& W''(x) \to +\, W''(x),   \qquad  A(t) \to  -\,A(t),
\end{eqnarray} 
leave the Lagrangian of our  theory, yet again, invariant (i.e. $L \to L$). Thus, we lay emphasis on the fact 
that we have parity as 
well as time-reversal symmetry  in the theory. The key difference between (18) and 
(20) is associated with the transformations of the fermionic variables $\psi$ and $\bar\psi$ 
under the time-reversal symmetry transformations $t \to - t$. Whereas in (18), there is
an $i$ factor in the transformations of the above variables, there is no $i$ factor
in the latter transformations. At the level of signatures, too, there is a difference
between the two transformations if we take a close look at them.

We wrap up this section with the remark that we have listed {\it only} three 
types of discrete symmetries in the above. However, in principle, there might be 
existence of more such symmetries. We shall see later, in our forthcoming sections, 
that the discrete symmetry transformation (20) would play a very crucial role in our 
present endeavour of establishing a connection between the symmetries and the cohomological operators
and it would also allow only SUSY potentials that are {\it even} under parity (i.e. $W(-x) = W(x)$). The latter correspond to the square integrable eigenfunctions for SUSY  QM \cite{2,3}.  
We further  note that, in all our three discrete symmetries (cf. (17), (18), (20)), the fermionic variable $\psi(t)$
transforms to $\bar \psi(t)$ and vice-versa. Thus, at present stage, these variables are `dual' to each-other.
We shall  see the consequences of this observation, later on,  when we shall focus more
on the presence of the duality transformations.

\section{Algebraic structures: cohomological aspects}

The continuous symmetry transformations $s_1, \; s_2$ and $s_w$ (cf. (11), (13)), 
in their operator form, obey the following algebraic structure:
\begin{eqnarray}
&&s^2_1 = 0, \qquad s^2_2 = 0, \qquad s_w = \{s_1, \; s_2\} = s_1\,s_2 + s_2\,s_1, \nonumber\\
&&[s_w, \; s_1] = 0, \qquad [s_w,\; s_2] = 0, \qquad s_w = (s_1 + s_2)^2.
\end{eqnarray}
This algebraic structure is exactly same as the algebra of de Rham cohomological operators 
$(d, \; \delta, \; \Delta)$ of differential geometry \cite{16,17,18} where $(\delta)d$ are the 
(co-)exterior derivatives
and $\Delta$ is the Laplacian operator. In explicit form, the algebra, satisfied by these 
cohomological operators, are \cite{16,17,18}
\begin{eqnarray}
&& d^2 = 0, \qquad \delta^2 = 0, \qquad \Delta = \{d, \; \delta\} = d\, \delta + \delta\,d, \nonumber\\
&& [\Delta, \; d] = 0, \qquad [\Delta,\; \delta] = 0, \qquad \Delta = (d + \delta)^2.
\end{eqnarray}
A close look at (21) and (22) tempts us to identify $(d, \, \delta, \, \Delta)$ with the 
set of continuous symmetries ($s_1, \, s_2, \, s_w$). However, there are other decisive  properties
that are {\it also} associated with  $(d, \, \delta, \, \Delta)$. These properties  have to be captured in the 
language of symmetry transformations of the Lagrangian if we wish to establish a {\it perfect} analogy between the 
cohomological operators and symmetries. For instance, first of all, we know that the nilpotent 
$(\delta^2 = d^2 = 0)$ (co-)exterior
derivatives $(\delta) d$  are connected with each-other by the Hodge duality ($*$)
operation defined on a given manifold. This important relationship is mathematically expressed as follows:  
\begin{eqnarray}
\delta = \pm\, * d\,*,\qquad \delta^2 = 0, \qquad d^2 = 0,
\end{eqnarray}
where $(\pm)$ signs are determined by the inner product of $p$-forms in a given 
dimension of the manifold without a boundary. For an even-dimensional manifold 
$\delta = -\, * d\,*$  and, for the odd-dimensional manifold, the signatures are decided 
by the degree of the forms, involved in the inner product, on that particular {\it odd}-dimensional manifold 
(see, e.g. \cite{16,17,18} for more details).

Within the realm  of theoretical physics, the relationships (23) are expressed in 
the language of symmetry properties \cite{19}. It can be seen that the 
interplay of the continuous and discrete symmetry transformations (cf. Sects. 3 and 4)  
leads to a relationship that is exactly similar in form as (23). For instance, we have the validity 
of the following relationship for a given generic variable $\Phi$ of the Lagrangian (10), namely;
\begin{eqnarray}
&& s_1\, \Phi = \pm\,* s_2 *\, \Phi, \qquad s_1^2 = s_2^2 = 0, \nonumber\\
&& \Phi = x,\, \psi,\, \bar \psi,\, A, \, W',\, W'',
\end{eqnarray}
where $*$ is the discrete symmetry transformations of Sect. 4. The $(\pm)$ 
signs in (23) are dictated by two successive operations of the 
discrete symmetry transformations  on the generic variable $\Phi$ (see, e.g. \cite{19} for details):
\begin{eqnarray}
*\, (*\, \Phi) = \pm\, \Phi, \quad\qquad \Phi = x,\, \psi,\, \bar \psi,\, A, \, W',\, W''.
\end{eqnarray}    
Furthermore, for our present model, there exists an  inverse relationship 
(i.e. $s_2\, \Phi = \mp\, * \,s_1 *\, \Phi$) corresponding to the relationship given in (24).

In view of the sacrosanct statements, made above about the duality-invariant theory, let us study the sanctity 
of the discrete symmetries (17), (18) and (20). In the case of (17), it can be checked that 
$* (*\, \Phi) = + \,\Phi$ for $\Phi = x,\, \psi,\, \bar \psi,\, A, \, W',$ $W''$. As it turns out, it can be 
explicitly verified that the relationships 
$s_1 \, \Phi = + \, *\, s_2 \,*\,\Phi$ (and/or $s_2\, \Phi = - \, *\, s_1\, * \, \Phi$) are
{\it not} satisfied. Physically, too, it is {\it not} allowed (see, e.g. \cite{2,3}) because
it corresponds to the superpotentials that are {\it odd} under parity (i.e. $W(-x) = - W(x)$).
  Similarly, in the case of (18), we check that $*(*\, \Phi_1) = + \,\Phi_1$
where $\Phi_1 = x, A, W^\prime, W^{\prime\prime}$ and $*(*\, \Phi_2) = - \,\Phi_2$
for $\Phi_2 = \psi,\, \bar \psi $. Here we have taken the generic 
variable $\Phi = (\Phi_1, \Phi_2)$. It is interesting to point out that the relations 
$s_1 \, \Phi_1 = + \, *\, s_2\, *\, \Phi_1$ (or its inverse  $s_2\, \Phi_1 = - \, *\, s_1\, * \, \Phi_1$) 
are {\it not} obeyed. In addition, the relations 
$s_1 \, \Phi_2 = - \, *\, s_2\,*\, \Phi_2$ (or its inverse  $s_2\, \Phi_2 = + \, *\, s_1\, * \, \Phi_2$) are 
also {\it not} satisfied for the transformations (18).
Thus, we shall discard both these sets of discrete symmetry transformations as they do 
not obey the sanctity of the strictures laid  down by the rules of a duality invariant physical theory
(see, e.g. \cite{19} for details).

Now let us concentrate on the discrete symmetry transformations (20) with the upper signature
(i.e. $x \to -\, x, \; t \to - \, t, \; \psi \to +\, \bar \psi,\; \bar \psi \to - \, \psi,\; 
A \to -\, A,\; W' \to - \, W', \; W'' \to + \,W''$). 
It  can be checked explicitly that
\begin{eqnarray}
&&*\, (* \,x) = +\, x, \qquad *\, (*\, \psi) = -\, \psi, \qquad * \,(*\, \bar \psi) = -\, \bar \psi, \nonumber\\
&&* \,(*\, A) = +\, A, \;\; *\, (* \,W') = +\, W', \;\; * \,(*\, W'') = +\, W''.
\end{eqnarray}
With these inputs, we readily  verify   (with $\Phi \equiv (\Phi_1,\; \Phi_2)$) that
\begin{eqnarray}
&&s_1\, \Phi_1 = + \, *\, s_2 *\, \Phi_1, \qquad s_2\, \Phi_1 = - \, *\, s_1 *\, \Phi_1, \nonumber\\
&&s_1\, \Phi_2 = - \, *\, s_2 *\, \Phi_2,  \qquad s_2\, \Phi_2 = + \, *\, s_1 *\, \Phi_2, \nonumber\\
&&  \Phi_1 = x,\; A, \; W',\; W'', \qquad \Phi_2 = \psi,\; \bar \psi.
\end{eqnarray}
Thus, we conclude that the discrete transformations (20), with the upper signature, are 
physically well-defined transformations that correspond to the Hodge duality $(*)$ 
operation of differential geometry as the relations (27) provide a physical realization 
of the relationship $\delta = \pm\, *\, d *$ between the (co-)exterior derivatives (of 
differential geometry) in terms of symmetries.

We close this section with the remark that the lower signature of the discrete 
symmetry transformations (20) {\it does not} lead to the correct relationships 
$ s_1\, \Phi_1 = +\, * \, s_2 \, * \Phi_1$ as well as $ s_1\, \Phi_2 = -\, * \, s_2 \, * \Phi_2$ 
(and their corresponding reverse relations). Thus, a {\it unique} set of discrete symmetry transformations
for our duality invariant physical theory, namely;
\begin{eqnarray}
&&x \to - \, x, \qquad t \to - \, t, \qquad \psi \to +\, \bar \psi, \qquad \bar \psi \to - \, \psi,\nonumber\\
&& A \to - \, A, \qquad W' \to - \, W', \qquad W'' \to + \, W'',
\end{eqnarray}  
is the one, we shall be concentrating on, for the rest of our discussions. 
First of all, we observe that under the above discrete symmetry transformations:
\begin{eqnarray}
*\, Q = +\,\bar Q, \quad *\,\bar Q = - \, Q, \quad *\, H = + \,H, \quad *\, L = +\,L,
\end{eqnarray}
where $*$ is nothing but the transformations (28). We point out that the transformations: 
$Q \to +\,\bar Q, \;  \bar Q \to -\, Q$ are analogous to the electromagnetic duality transformations  
for the source-free Maxwell equations where ${\bf E} \to +\, {\bf B}$ and ${\bf B} \to -\, {\bf E}$. 
This is the reason that, at the end of Sect. 4, we have claimed that the variables $\psi$ and $\bar \psi$ are 
``dual" to each-other because these variables distinguish the two SUSY nilpotent and conserved 
charges ($Q, \bar Q$). The second point to be noted is the fact that the transformations (28) 
allow superpotentials that are {\it even} under parity (i.e. $W(-x) = W(x)$) and these are 
the ones that are physically interesting (see, e.g. \cite{2,3}).

\section{Conserved charges and cohomological operators: mappings}

There is still {\it one} issue which has {\it not yet} been settled as far as the perfect 
analogy between the de Rham
cohomological operators $(d,\; \delta,\; \Delta)$ and the conserved charges $(Q, \; \bar Q,\; H)$ is
concerned. We know that the operation of $d$ on a differential form $(f_n)$ of degree $n$ raises 
it to a form  $(f_{n+1})$ of degree $(n + 1)$ (i.e. $d f_n \sim f_{n+1}$) where $n = 0, 1, 2,...$
On the contrary, the action of the co-exterior derivative on a form $(f_n)$ of degree $n$ lowers the
degree of the form by one (i.e. $\delta f_n \sim f_{n-1}$) where $n = 1, 2, 3,...$  Due to 
$\Delta = (d + \delta)^2 \equiv d\, \delta + \delta\, d$, it is clear that the Laplacian operator $\Delta$ does not 
affect the degree of a form on which it operates (i.e. $\Delta f_n \sim f_n$) (see. e.g. \cite{16,17,18}).

To capture the above properties, in the language of symmetry generators, 
we note the following straightforward algebra:
\begin{eqnarray}
&&\big[Q\,\bar Q, \; Q\big] = + \,Q\, H, \qquad \big[Q\,\bar Q, \; \bar Q\big] = - \,H\,\bar Q,\nonumber\\
&&\big[\bar Q\,Q, \; Q\big] = -\,H\,Q, \qquad \big[\bar Q\,Q, \;  \bar Q\big] = +\,\bar Q\, H.
\end{eqnarray} 
It is to be emphasized that we have normalized the expressions for the supercharges
$Q,\, \bar Q$ with some constant factors so that the basic SUSY algebra: $Q^2 = \bar Q^2 = 0, \, \{Q, \, \bar Q\}
= H,\, [H,\, Q] = [H,\, \bar Q] = 0$ is satisfied.\footnote{The details of these aspects of 
the algebraic structures  could be found in our Appendix A where the charges have been redefined suitably.}  
We point out the fact that $\dot Q = -\,i\, [Q,\, H] = 0$ and 
$\dot {\bar Q} = -\, i\, [\bar Q,\, H] = 0$
imply that $H\, Q = Q\, H, \, H\, \bar Q = \bar Q\, H$. The latter relations lead to:
$H^{-1}\,Q = Q\, H^{-1},$ $H^{-1}\,\bar Q = \bar Q\, H^{-1}$ if the inverse of the Hamiltonian exists.
Since we are focusing on the non-singular Hamiltonian (in the matrix form), we presume that the  Casimir operator $H$
has its well-defined inverse. The latter, too, would be the Casimir operator for the whole   algebra
(i,e.  $(H^{-1},\, Q,\, \bar Q)$). As a consequence, we have an algebraically  suitable  form of (30), as follows
\begin{eqnarray}
\left[\frac{Q\, \bar Q}{H},\; Q\right] = +\, Q, \qquad 
\left[\frac{Q\, \bar Q}{H},\; \bar Q\right] = -\, \bar Q,\nonumber\\
\left[\frac{\bar Q\, Q}{H},\; Q\right] = -\, Q, \qquad 
\left[\frac{\bar Q\, Q}{H},\; \bar Q\right] = + \, \bar Q.
\end{eqnarray} 
These relations would play very important roles in our further discussions as would become clear
when we shall discuss about the ladder operators in the language of the conserved charges.

Let us now define an eigenstate $|\xi \rangle_p$ w.r.t. the operator $\left(Q\, \bar Q/H\right)$
(i.e. $\left(Q\, \bar Q/H\right) |\xi\rangle_p = p\, |\xi\rangle_p$) where $p$ is the eigenvalue. By exploiting the 
algebra (31), it is straightforward to check that
\begin{eqnarray}
&&\left(\frac{Q\, \bar Q}{H}\right) Q \;|\xi\rangle_p = (p + 1) \, Q \;|\xi\rangle_p,\nonumber\\
&&\left(\frac{Q\, \bar Q}{H}\right) \bar Q\; |\xi\rangle_p = (p - 1) \,\bar Q \;|\xi\rangle_p,\nonumber\\
&&\left(\frac{Q\, \bar Q}{H}\right) H \;|\xi\rangle_p = p \; Q \;|\xi\rangle_p.
\end{eqnarray}
The above equation establishes the fact that the states $Q\, |\xi\rangle_p,$ $\bar Q \,|\xi\rangle_p$
and $H\, |\xi\rangle_p$ have the eigenvalues $(p + 1), \; (p - 1), \;p$, respectively. Thus, if we identify 
the eigenvalue $p$ with the degree of a form, then, it is clear that the following mapping between the 
conserved charges and cohomological operators of differential geometry emerge, namely;
\begin{eqnarray}
\big(Q, \; \bar Q,\; H\big) \longleftrightarrow \big(d, \; \delta,\; \Delta\big).
\end{eqnarray}
We point out that the property of (lowering)raising of a given  differential form by the 
operations of the (co-)exterior derivatives is also captured in the language of the
conserved charges when we identify the eigenvalue of the operator $\left(Q\, \bar Q/H\right)$  
with the degree of a given differential form.

There is yet another representation of $\big(d, \; \delta,\; \Delta\big)$ in the language of the 
eigenvalues of a set of conserved charges $\big(Q, \; \bar Q,\; H\big)$. To this end in mind, 
we define an eigenstate  $|\chi\rangle_q$, with the eigenvalue $q$, as:
\begin{eqnarray}
\left(\frac{\bar Q\, Q}{H}\right) |\chi\rangle_q = q \;|\chi\rangle_q.
\end{eqnarray}
Exploiting the structure of (31), it is straightforward to verify the following 
consequences that ensue due to (34), namely;
\begin{eqnarray}
&&\left(\frac{\bar Q\, Q}{H}\right) Q \;|\chi\rangle_q = (q - 1) \, Q\; |\chi\rangle_q,\nonumber\\
&&\left(\frac{\bar Q\, Q}{H}\right) \bar Q\; |\chi\rangle_q = (q + 1) \,\bar Q\; |\chi\rangle_q,\nonumber\\
&&\left(\frac{\bar Q\, Q}{H}\right) H \;|\chi\rangle_q = q \, Q\; |\chi\rangle_q,
\end{eqnarray}
which establishes that the states $\bar Q \,|\chi\rangle_q,\;  Q\, |\chi\rangle_q$
and $H\, |\chi\rangle_q$ have the eigenvalues $(q + 1), \; (q - 1), \;q$, respectively.
Thus, as far as the lowering and raising property of $\big(d, \; \delta,\; \Delta\big)$ is concerned,
we have the following mapping between the conserved charges $\big(Q, \; \bar Q,\; H\big)$ and 
the cohomological operators:
\begin{eqnarray}
\big(\bar Q, \; Q,\; H\big) \longleftrightarrow \big(d, \; \delta,\; \Delta\big).
\end{eqnarray}
Thus, ultimately, we conclude that if the degree of a given form is identified with the 
eigenvalue of a given state (in the quantum Hilbert space) corresponding to the operator 
$\left(\bar Q\, Q/H\right)$, then, the consequences ensuing from the
 operation  of the set of charges $\big(\bar Q, \;  Q,\; H\big)$  is
identical to the operation of the set $\big(d, \; \delta,\; \Delta\big)$ on the degree 
of a given form. This is why the mapping (36) is correct. We close this section with 
the final remark that we have provided two physical realizations of the cohomological 
operators in terms of the conserved charges of the general ${\cal N} = 2$
SUSY quantum mechanical model at the algebraic level.

\section{Conclusions}
In our present investigation, we have given a concrete proof of our earlier conjecture that any 
arbitrary ${\cal N}= 2$ SUSY quantum mechanical model would provide a cute physical example of the Hodge theory. 
We have derived the general Lagrangian for the ${\cal N} = 2$ SUSY quantum mechanical theory by 
exploiting the basic tenets of (i) the ${\cal N} = 2$ SUSY supervariable and its expansion along the 
Grassmannian directions of a $(1, 2)$-dimensional supermanifold (ii) the idea of supercovariant 
derivatives, and (iii) the Taylor expansion of  an arbitrary superpotential (cf. (8)). 
We have also demonstrated that the discrete as well as continuous 
symmetry properties of this Lagrangian (and their corresponding generators) provide the 
physical realizations of the cohomological operators. As a consequence, the general ${\cal N}=2$ 
SUSY model is a physical example of the Hodge theory because all the cohomological operators 
find their physical realizations in the language of the interplay between the underlying 
discrete as well as continuous symmetry transformations of our present theory.

In addition to the well-known continuous symmetry transformations, generated by the conserved 
charges $\big(Q, \;  \bar Q,\; H\big)$, we have discussed various kinds of discrete symmetries 
in our present endeavour (cf. (17), (18), (20)). Out of these {\it three} discrete symmetry transformations, 
only {\it one} (i.e. (20)) is the perfect symmetry for a duality-invariant theory.\footnote{In fact, there 
are two discrete symmetries in (20). Both of them are {\it not} appropriate. Out of these two, only one is 
physically correct for a ``duality'' invariant theory \cite{19} which is given in (28).} We have shown that, 
under the perfect discrete symmetry transformations (20), the supercharges $Q$ and $\bar Q$ 
transform in exactly the same manner 
as the duality transformations for the electric and magnetic fields of a source-free Maxwell equations. Further, the 
above unique discrete symmetry transformation turns out to provide a physical realization of the 
Hodge duality operation of differential geometry. Thus, 
for our present general ${\cal N} = 2$ SUSY model, we have also been able to 
provide the physical realizations of the relationships (i.e. $\delta = \pm\, *\, d\,*$)
 between the nilpotent ($d^2 = \delta^2 = 0$) of order two (co-)exterior derivatives ($\delta(d)$) of the 
differential geometry in the language of symmetries.

As pointed out, in the concluding remark at the fag end of Sect. 4, there might exist
many discrete symmetries in the theory, under which, the Lagrangian (10) would remain
invariant. The decisive feature of a {\it physically} relevant discrete symmetry would always be, however,
the validity of the relations $s_1 \Phi = \pm * s_2 * \Phi, s_2 \Phi = \mp * s_1 * \Phi $
for  a generic variable  $\Phi$ where $(\pm)$ signs would be dictated by the signatures
of $*\, (*\, \Phi) = \pm \, \Phi$. In this context, we note that, under the 
following two sets of discrete symmetry transformations
\begin{eqnarray}
&& x \to - x, \qquad t \to + t, \qquad \psi \to + \bar\psi, \qquad \bar \psi \to + \psi, \nonumber\\
&& A(t) \to + A(t), \quad W^\prime (x) \to + W^\prime (x), \nonumber\\
&& W^{\prime\prime} (x) \to - W^{\prime\prime},
\end{eqnarray}
\begin{eqnarray}
&& x \to + x, \qquad t \to - t, \qquad \psi \to \pm \bar\psi, \qquad \bar \psi \to \mp  \psi, \nonumber\\
&& A(t) \to + A(t), \quad W^\prime (x) \to + W^\prime (x), \nonumber\\
&& W^{\prime\prime} (x) \to + W^{\prime\prime},
\end{eqnarray}
the Lagrangian (10) remains invariant (i.e. $L \to L$). However, these do {\it not} satisfy the 
sacrosanct relations  $s_1 \Phi = \pm * s_2 * \Phi,\; s_2 \Phi = \mp * s_1 * \Phi $. In addition, 
the above discrete symmetry transformations imply that the superpotentials must be {\it odd} under parity 
(i.e. $W(-x) = - W(x)$). However, such kind of potentials are {\it not} allowed by the SUSY QM as they do 
not always lead to the square integrable eigenfunctions. Hence, the above sets of discrete symmetry 
transformations are {\it not} physically interesting at all to us. Furthermore, the correct transformations 
of $Q$ and $\bar Q$, under the discrete symmetry transformations, also shed light on the correctness
of a chosen set of discrete symmetry transformations for our present SUSY quantum mechanical theory.

It would be very nice endeavour to exploit our present observations in the context of the study of
 ${\cal N} = 2$ SUSY (non-)Abelian gauge theories (in any arbitrary dimension of spacetime)
where there is a possibility of the appearance of 
cohomological structure. Such theories might be shown to be the perfect models for the Hodge theory 
as well as new models for TFTs. The latter are expected to turn out to be 
different from the Witten-type TFTs \cite{12} as  well as Schwarz-type TFTs \cite{13}. We have performed
such kind of study in the case of {\it usual} 2D (non-)Abelian gauge theories (see, e.g. \cite{6,11}) where the 
full strength of the symmetries of the Hodge theory has been exploited in its full blaze of glory. 
Our work can be possibly extended to contain exotic discrete symmetries that have been mentioned in \cite{21,22}.
This exercise will definitely enrich the mathematical structure of our present analysis where
the existence of discrete symmetries  plays  an important role.
Furthermore, we plan to explore the possibility of existence of the cohomological structure 
in the ${\cal N} = 4$ SUSY quantum mechanical theories \cite{23}. In fact, we expect many possibilities
of the physical realizations of the cohomological operators in this case. Our present work
can also be extended to higher dimensional (e.g. 2D and 3D) SUSY quantum mechanical models
following the work done in \cite{24}. We are devoting time on the above 
cited problems  and our results would be reported, later on, in our future publications \cite{25}.

\begin{acknowledgements}
RK would like to express his 
deep sense of gratitude to the UGC, Government of India, for the financial   
support through SRF scheme.
\end{acknowledgements}

\appendix
\section{On the derivation of ${\cal N} = 2$ SUSY algebra}

The whole of algebraic structures in Sect. 6 are based on the basic ${\cal N} = 2$ SUSY algebra 
$Q^2 = \bar Q^2 = 0, \; \{Q, \; \bar Q\} = H, \; [H, \; Q] = [H, \;\bar Q] = 0$ which 
is satisfied on the {\it on-shell}. To corroborate this statement, first of all, auxiliary variable $A$ in the 
Lagrangian (10) is replaced by $(- W')$ due to the equation of motion $A = - W^\prime$
(which emerges from (10) itself). Furthermore,
 the symmetry transformations $s_1$ and $s_2$ 
(cf. (11)) are modified a bit by the overall constant factors. Thus, we have the following 
different looking Lagrangian
\begin{eqnarray}
L_0 = \frac{1}{2}\, {\dot x}^2 + i\,\bar \psi \, \dot \psi - \frac {1}{2}\, (W')^2 
+ W'' \, \bar \psi\, \psi, 
\end{eqnarray}
which remains invariant under the following transformations
\begin{eqnarray}
&&s_1 x = -\frac{1}{\sqrt 2}\,i\, \psi, \quad s_1 \bar \psi = \frac{1}{\sqrt 2}\,(\dot x - i\,W'), 
\quad s_1  \psi = 0, \nonumber\\
&&s_2 x = + \frac{1}{\sqrt 2}\,i\, \bar \psi, \quad s_2 \psi = - \frac{1}{\sqrt 2} (\dot x + i\,W'), 
\;\;\, s_2 \bar \psi = 0. 
\end{eqnarray}
The above transformations are nilpotent of order two (i.e. $s_1^2 = s_2^2 = 0$) only when 
the equations of motion $\dot \psi - i\, W''\, \psi = 0, \; \dot {\bar \psi} + i\, W''\, \bar \psi = 0 $ 
are used. It can be checked that $s_1 L_0 = d/dt  \left(W^\prime \psi/ \sqrt 2\right)$ and 
$s_2 L_0 = d/dt  \left(i \bar \psi \dot x/ \sqrt 2\right)$. Hence, the action integral $S = \int dt L_0$
remains invariant under $s_1$ and $s_2$.

The conserved Noether charges, that emerge corresponding to (A.2), are
\begin{eqnarray}
Q = -\,\frac{1}{\sqrt 2}\,\big(i\,\dot x + W'\big)\, \psi, \qquad 
\bar Q = +\,\frac{1}{\sqrt 2}\, \bar \psi\,\big(i\,\dot x - W'\big).
\end{eqnarray}
These charges are same as quoted in (15) except the fact that $A$ has been replaced by $(- W')$ 
(due to the equation of motion from the Lagrangian (10)) and the constant factors $(\mp 1/\sqrt 2)$ 
have been included for the algebraic convenience.  
It can be readily checked that the above charges are the generator for the transformations (A.2) 
because  we have the following relationships:
\begin{eqnarray}
s_1 \Phi = \pm\,i\, \big[\Phi, \; Q\big]_\pm, \qquad \quad 
s_2 \Phi = \pm\,i\, \big[\Phi, \; \bar Q\big]_\pm, 
\end{eqnarray}
where the generic variable $\Phi$ corresponds to the variables $x,\, \psi, \, \bar \psi$ and the subscripts $(\pm)$ on 
square  brackets stand for the (anti)commutator depending on the generic variable $\Phi$
being (fermionic)bosonic in nature. The $(\pm)$ signs, in front of the brackets, are {\it also} chosen judiciously
(see, e.g. \cite{20} for details).

The structure of the specific ${\cal N} = 2$ SUSY algebra now follows when we exploit the basic relationship 
(A.4). In other words, we observe the following
\begin{eqnarray}
&& s_1 Q = i \,\big\{Q, \; Q\big\} = 0 \Longrightarrow  Q^2 = 0,\nonumber\\ 
&& s_1 \bar Q = i\, \big\{\bar Q, \; \bar Q\big\} = 0 \Longrightarrow \bar Q^2 = 0,\nonumber\\ 
&& s_1 \bar Q = i\, \big\{\bar Q, \; Q\big\} = \,i\,H \Longrightarrow \big\{\bar Q, \;  Q\big\} = \, H,\nonumber\\ 
&& s_2 Q = i\, \big\{Q, \; \bar Q\big\} = \,i\,H \Longrightarrow  \big\{Q, \; \bar Q\big\} = \,H, 
\end{eqnarray} 
where $H$ is the Hamiltonian (corresponding to the Lagrangian (A.1)). The explicit form of $H$ 
can be mathematically expressed as:
\begin{eqnarray} 
H &=& \frac{1}{2}\, {\dot x}^2 + \frac {1}{2}\, (W')^2 
- W'' \, \bar \psi\, \psi \nonumber\\
& \equiv & \frac{1}{2}\, p^2 + \frac {1}{2}\, (W')^2 
- W'' \, \bar \psi\, \psi, 
\end{eqnarray}
where $p = \dot x$ is the momentum corresponding to the variable $x$. We also lay emphasis 
on the fact that we have exploited the inputs from  equations of motion 
$\dot \psi - i\, W''\, \psi = 0, \, \dot {\bar \psi} + i\, W''\, \bar \psi = 0 $  
in the derivation of $H$ from the Legendre transformation 
$H = \dot x\, p + \dot \psi \, \Pi_\psi + \dot{\bar \psi}\, \Pi_{\bar \psi} - L$ where
$\Pi_\psi = - i\, \bar \psi$ and $\Pi_{\bar\psi} = 0$.
The derivation of specific
${\cal N} = 2$ SUSY algebra (cf. (A.5)) is very straightforward because we have used {\it only} 
(A.2) and (A.3) in the calculation of l.h.s. of (A.5) from  which, the results of the r.h.s. 
(i.e. specific ${\cal N} = 2$ SUSY algebra) trivially ensue.

We wrap up this Appendix with the remarks that the specific ${\cal N} = 2$ SUSY algebra
$Q^2 = \bar Q^2 = 0, \; \{Q, \; \bar Q\} = H$, listed in (A.5), is valid {\it only} on the on-shell 
where the validity of the Euler-Lagrange equations of motion is taken into account. 
Furthermore, it may be trivially noted that, for the choices 
$W' = \omega\, x$ and $W' = \omega\, f(x)$ in the Lagrangian (A.1), we obtain the Lagrangians 
for the SUSY harmonic oscillator and its generalization in \cite{5}. For the description of the motion 
of a charged particle in the $X-Y$ plane under the influence of a magnetic field along $Z$-direction, 
the choice for $W'$ could  be found  in the standard books on SUSY quantum mechanics and relevant literature 
(see, e.g. \cite{2,3}).

\bibliographystyle{model1a-num-names}

\end{document}